\documentclass[sigconf]{acmart}

\pdfoutput=1
\usepackage{booktabs} 
\usepackage{multirow}
\usepackage{enumitem}
\usepackage{balance}
\usepackage[ruled]{algorithm2e}
\usepackage[flushleft]{threeparttable}
\usepackage{bm}
\usepackage{xcolor}

\setlist[itemize]{leftmargin=*}
\setcopyright{acmcopyright}

\acmDOI{10.1145/1122445.1122456}
\acmISBN{978-1-4503-XXXX-X/18/06}

\acmYear{2018}
\copyrightyear{2018}

\acmArticle{4}
\acmPrice{15.00}

\begin{document}

\title{Context-aware Reranking with Utility Maximization for Recommendation}
\author{Yunjia Xi$^1$, Weiwen Liu$^2$, Xinyi Dai$^1$, Ruiming Tang$^2$, Weinan Zhang$^1$}
\author{Qing Liu$^2$, Xiuqiang He$^2$, Yong Yu$^1$}
\affiliation{$^1$Shanghai Jiao Tong University, $^2$Huawei Noah's Ark Lab}
\email{{xiyunjia, daixinyi, wnzhang, yyu}@sjtu.edu.cn} \email{{liuweiwen8, tangruiming,liuqing48,hexiuqiang1}@huawei.com}

\newcommand{\xinyi}[1]{{\color{cyan} #1}}

\renewcommand{\shortauthors}{Y. Xi, W. Liu, X. Dai et al.}

\settopmatter{printacmref=false}

\begin{abstract}
This paper provides a sample of a \LaTeX document which conforms,
somewhat loosely, to the formatting guidelines for
ACM SIG Proceedings.\footnote{This is an abstract footnote}
\end{abstract}


\begin{abstract}
As a critical task for large-scale commercial recommender systems, reranking has shown the potential of improving recommendation results by uncovering mutual influence among items.  Reranking rearranges items in the initial ranking lists from the previous ranking stage to better meet users' demands. 
However, rather than considering the context of initial lists as most existing methods do, an ideal reranking algorithm should consider the \textit{counterfactual context} -- the position and the alignment of the items in the \textit{reranked lists}. In this work, we propose a novel pairwise reranking framework, Context-aware Reranking with Utility Maximization for recommendation (CRUM), which maximizes the overall utility after reranking efficiently. Specifically, we first design a utility-oriented evaluator, which applies Bi-LSTM and graph attention mechanism to estimate the listwise utility via the \textit{counterfactual context} modeling. Then, under the guidance of the evaluator, we propose a pairwise reranker model to find the most suitable position for each item by swapping misplaced item pairs. Extensive experiments on two benchmark datasets and a proprietary real-world dataset demonstrate that CRUM significantly outperforms the state-of-the-art models in terms of both relevance-based metrics and utility-based metrics. 
\end{abstract}


\maketitle

\section{Introduction}

Recommender System (RS) has been widely deployed in websites and mobile applications, including e-commerce \cite{youtube, zhou18kdd}, videos \cite{youtube, netflix}, and news \cite{Das2007GoogleNP, newRes}. A commercial RS consists of three stages in general, i.e., candidate generation, ranking, and reranking.
Thousands of relevant candidates are surfaced in the candidate generation stage, followed by a ranking function to score and select top items in the ranking stage. The last reranking stage further rearranges items in the initial ranking lists from the previous stage by attending to mutual influence between items, the results of which directly affect user satisfaction as well as the revenue of the RS. Recent attention is increasingly focused on the reranking stage due to its desired outcomes  \cite{dlcm, prm, setrank, irgpr}.  

Foundational work in reranking has shown the potential of improving recommendation results by uncovering mutual influence among items \cite{seq2slate, dlcm, prm, irgpr}. In fact, how likely a user favors an item is affected by other items placing in the same list. As such, existing reranking algorithms have focused primarily on designing sophisticated models like recurrent neural networks (RNN) \cite{seq2slate, dlcm}, Transformer \cite{prm, setrank}, or graph neural networks (GNN) \cite{irgpr} to extract such mutual influences based on the context of initial ranking lists. However, an ideal reranking algorithm, we posit, should be aware of the current item's position and the alignment of other items in the \textit{reranked lists} -- referred to as the \textit{counterfactual context} -- rather than the input initial lists. The counterfactual context describes where we permutate the initial ranking lists and obtain the lists that have never been displayed to users by the system.


\begin{figure}[t]
	\centering
	\includegraphics[width=0.9\columnwidth,trim=60 40 50 0,clip]{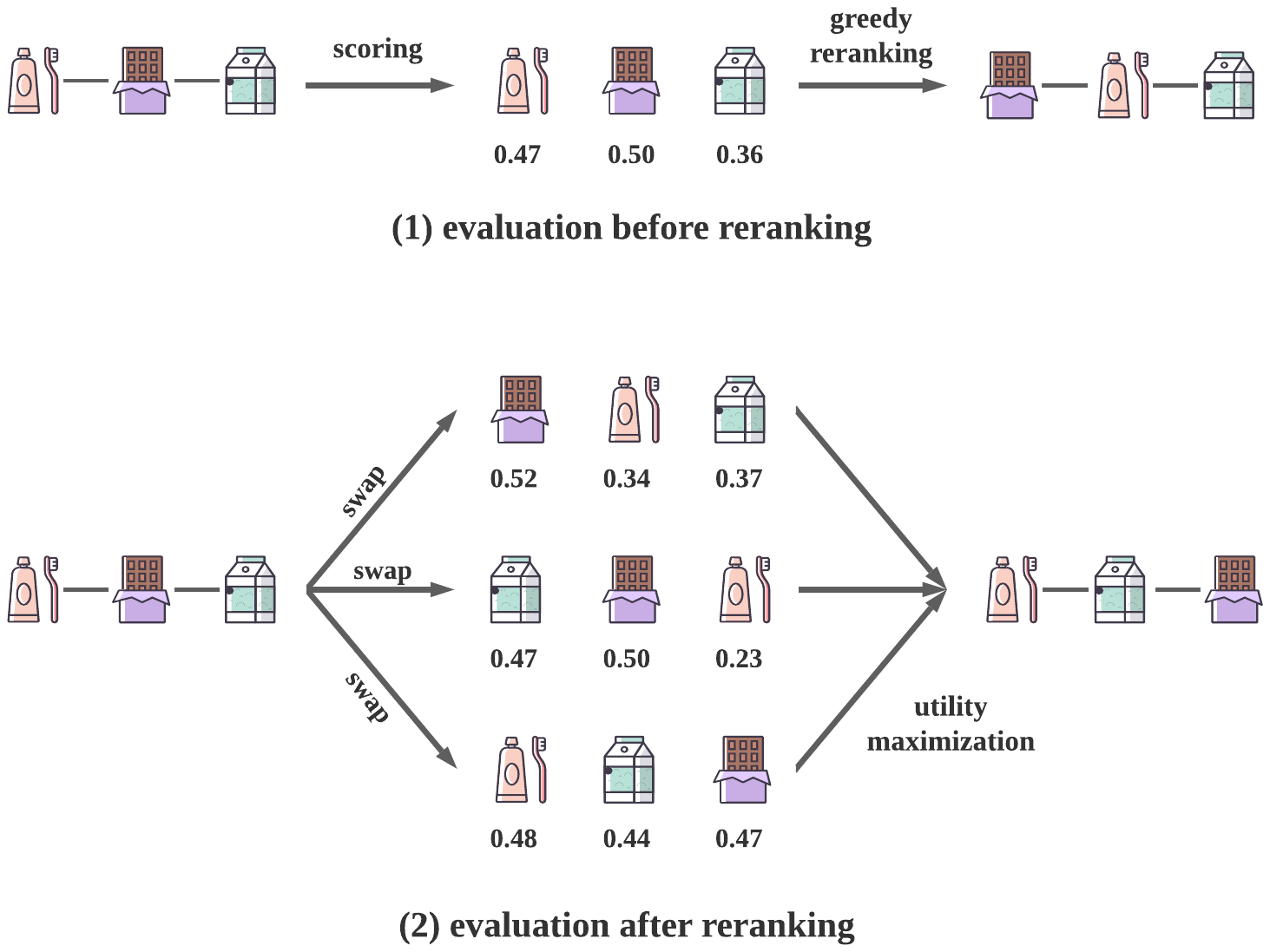}
	\caption{Comparison between evaluation before and after reranking methods.}
	\vspace{-10pt}
	\label{fig:comparison}
\end{figure}
Therefore, we are interested in learning from counterfactual context for reranking.  The user's listwise utility (e.g., total clicks or revenue) is influenced not only by the relevance of items but also by the locations and surrounding items \cite{joachims2005, lorigo2008eye, YZ2017, LORIGO20061123}, so that even with the same set of items, the listwise utility varies with different permutations. Yet current works estimate the utility simply based on the context of the initial lists and then rerank the items by their estimated scores following a greedy strategy, ignoring the fact that such reranking operation already modifies the actual utility. We refer to this type of reranking model as the \textit{evaluation-before-reranking method}. On the one hand, the utility is sensitive to positions. Evaluation-before-reranking methods, however, only use the initial position information, leading to its failure in modeling the item's utility in other positions. On the other hand, the reranking operation usually changes the neighboring items, bringing different mutual influences between items from the original list.

As shown in Figure \ref{fig:comparison}, for an initial ranking list (dental cares, chocolate, milk), the evaluation-before-reranking methods predict the utility directly under the context of such an initial list and the estimated value is $(0.47, 0.5, 0.36)$. Then the items are greedily reranked. However, $0.47$ represents the utility when dental cares are placed at position $1$, before chocolate and milk. Once we place dental cares at position $2$ according to the descending order of estimated utility, the context of dental cares is modified, which makes the previous estimation imprecise.

To this end, we aim to maximize the overall utility after reranking by finding the optimal permutation with different counterfactual contexts. Such an objective indicates that we cannot adopt an evaluation-before-reranking approach but an \textit{evaluation-after-reranking approach}. Figure \ref{fig:comparison} presents an example. The evaluation-after-reranking approach first generates feasible candidate reranked lists, e.g., by swapping two items in the initial list. Then the utility is evaluated according to the current counterfactual context, and the list with maximal overall utility is chosen. 

However, this brings the following two new challenges to the reranking stage:
1) \textbf{Evaluation after reranking.} Users' feedback on counterfactual context is unobtainable, as it is impractical to ask users to provide feedback for every permutation of a given list. Therefore, the key concern is how to estimate the utility of the reranked lists precisely according to the difference between observed context and counterfactual context. 
2) \textbf{Exponential time complexity.} Directly finding the optimal reranking list with counterfactual context is a combinatorial optimization problem and has $n!$ feasible permutations. Generating and evaluating all possible permutations at the inference stage is computationally expensive.

To resolve the aforementioned issues, we propose a pairwise reranking framework, Context-aware Reranking with Utility Maximization for recommendation (CRUM), which consists of a position-aware graph embedding, a utility-oriented evaluator, and a reranker. Firstly, to capture the interaction information between items and positions, we construct a novel positional graph and extract the position-aware graph embedding. Next, the utility-oriented evaluator evaluates the new reranked list by modeling the sequential browsing of users and outputs the listwise utility to solve the challenge of \textit{evaluation after reranking}. Lastly, the reranker derives feasible reranked lists by swapping pairs of items with the guidance of the evaluator. Instead of generating all the permutations, we only adjust the mismatched pairs towards the optimal ranking. We compare the utility given by the evaluator before and after the swap of two items, and update the model with an efficient Lambdaloss framework. Lists are reranked via a well-designed scoring function, which reduces the computation complexity from $O(n!)$ to $O(n)$ at the inference stage, and thus solves the challenge of \textit{exponential time complexity}.


To summarize, the contributions of our work are as follows:
\begin{itemize}
    \item We highlight the necessity of leveraging counterfactual context to evaluate the listwise utility after reranking for more precise estimation. We develop a general evaluation-after-reranking solution to learn the reranking strategy for optimizing the overall utility.
    \item To avoid exponential solutions, we propose a novel utility-oriented reranking framework, CRUM, with the position-aware graph embedding to extract mutual influence between items and positions, an evaluator to estimate the listwise utility using counterfactual context, and a pairwise reranker to find the most suitable position after reranking for items in a perspective of swap.
    \item Extensive experiments are conducted on two widely-used public datasets and a proprietary real-world  recommendation dataset. Those experiments demonstrate the effectiveness of CRUM, which outperforms the state-of-the-art models w.r.t both relevance-based metrics like MAP and utility-based metrics like CTR. 
\end{itemize}


\section{Related Work}

\subsection{Learning to Rank}
The reranking stage is built on initial rankings given by the ranking stage. Learning to rank that applies machine learning algorithms is one of the most widely used methods in ranking stage.
According to the loss function, it can be broadly classified into pointwise \cite{mcrank,GBM,prank}, pairwise \cite{svmrank, ltrGD, GBRank}, and listwise \cite{burges2010ranknet, cao2007, softrank, adarank, ListMLE} methods. The pointwise methods, e.g., McRank \cite{mcrank} and PRank \cite{prank}, regard ranking as a classification or regression problem and predict an item's relevance score
at a time. The pairwise methods like SVMRank
convert the ranking to a pairwise classification problem to optimize the relative positions of two items. 
The listwise methods directly maximize ranking metrics of lists. For example, LambdaMART \cite{burges2010ranknet} combines boost tree model MART \cite{sgb,GBM} and LambdaRank \cite{lambdarank} to optimize NDCG directly.

In the experiment part of this paper, we discuss how different types of learning-to-rank methods affect the performance of the reranking models.

\begin{figure*}[t]
	\centering
	\includegraphics[width=0.95\textwidth,trim=310 490 290 100,clip]{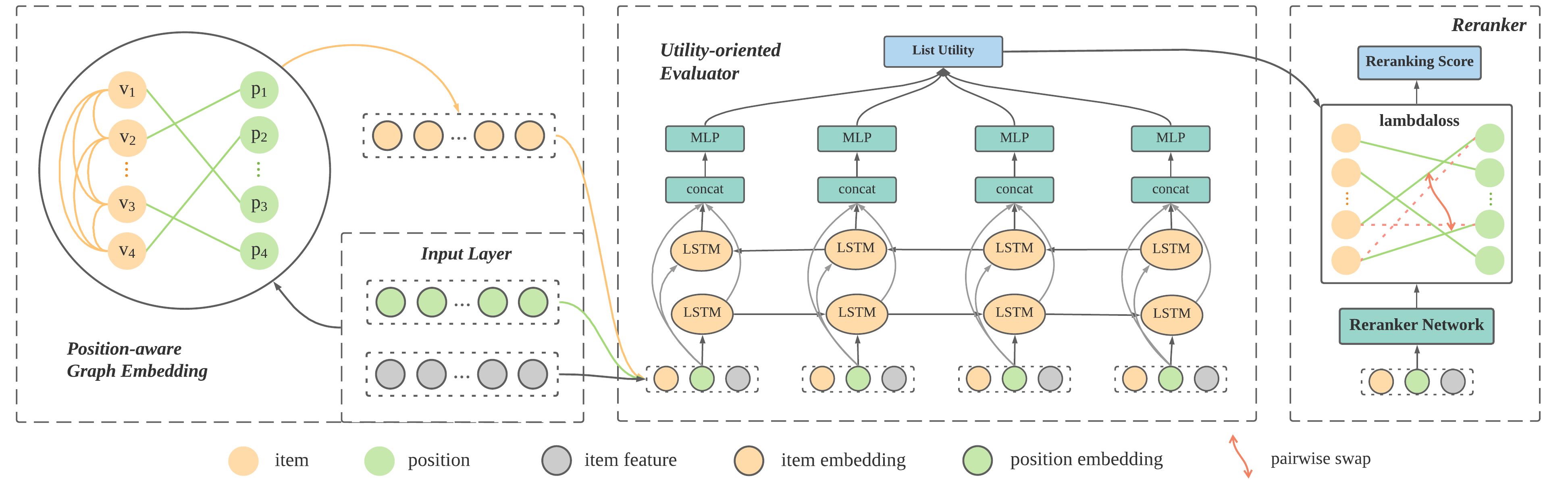}
	\vspace{-5pt}
	\caption{The overall framework of URCC.
	}
	\vspace{-5pt}
	\label{fig:framework}
\end{figure*}

\subsection{Reranking}
Compared to ranking methods, reranking methods utilize initial lists
and emphasize the mutual influence between items. Depending on how to model the mutual influence of items, these methods can be roughly divided into three categories: RNN-based \cite{seq2slate, dlcm, feng2021revisit}, Transformer-based \cite{prm, setrank}, and GNN-based \cite{irgpr} methods. 

RNN-based methods apply RNN to model the item interaction and implicitly extract positional information from the initial ranking. DLCM \cite{dlcm} applies GRU to encode the whole ranking list into the representation of items. Seq2Slate \cite{seq2slate} uses pointer network with a decoder to directly generates the ranked list. PRS \cite{feng2021revisit} consists of PMatch and PRank, where PMatch obtains the candidate lists and PRank adopts Bi-LSTM to evaluate lists. However, it is heuristic without a learnable scoring function. Moreover, its high computational complexity for online inference limits its applications.
Due to Transformer's ability to model the interaction between any two items in $O(1)$ distance, PRM \cite{prm} adopts it to encode the mutual influences between items. SetRank \cite{setrank} employs multi-head self-attention blocks to capture the local context information and permutation-equivalent representations of items. 
Usually, position embedding or a designed position embedding function is employed in these Transformer-based models.
The GNN-based model IRGPR \cite{irgpr} explicitly models item relationships by recursively aggregating relational information from multi-hop neighborhoods. Though this work makes advances, the required additional item relationship restricts its usage to specific applications like e-commerce. 

To summarize, above models employ sophisticated mechanisms to model the mutual influence between items. Whereas most of them belong to evaluation-before-reranking approaches.
The only exception, PRS \cite{feng2021revisit}, is heuristic and has relatively high complexity. Unlike previous works, our model not only considers counterfactual context, but also learns a scoring function to reduce complexity.

 \section{Problem Formulation}\label{section:preliminary}

 
  A reranking model generally aims at generating better ranking lists by using initial ranking lists arranged by the previous ranker. Given a user's request $r \in \mathcal{R}$, the initial ranker returns an initial ranked list $L_r$ with $n$ items. Then, the reranking model reranks the initial list and generates a list that better meets the user's needs. Mathematically, We denote list interaction logs that the reranking model uses for the request $r$  as $\{(\mathbf{x}_{i}, \gamma_{i}, k_{i}, c_{i,k_{i}})_{i\in L_r}\}$, where each item $i$ in the ranking list $L_r$ is associated with a feature vector $\mathbf{x}_{i}\in\mathbb R^{d_x}$ of size $d_x$ and a utility value $\gamma_{i}\in\mathbb R$ (e.g., $\gamma_{i}$ can be the bid price of each ad in sponsored search). We denote the initial position of item $i$ as $k_i$ 
  and $c_{i, k_{i}}$ is the user's implicit feedback on item $i$ at position $k_{i}$, i.e., $c_{i, k_{i}}=1$ for click and $c_{i, k_{i}}=0$ for non-click. 

  
 
The target of reranking is to present better item arrangements than those produced by the initial ranker. We regard better item arrangements as arrangements that yield more utility, and define the utility of request $r$ as the expected sum of the weighted click of each item in the ranking list $L_r$ under a specific permutation $\pi$,
 \begin{equation}
\begin{aligned}
     U_{ \pi}(r) = \mathbb{E} \Big[\sum_{i\in L_r} c_{i, \pi(i)} \cdot \gamma_{i} \Big]
     = \sum_{i\in L_r} P(c_{i, \pi(i)} = 1) \cdot \gamma_{i} ~,
\end{aligned}
\label{eq:utility}
\end{equation} 
where $\pi$ is a feasible permutation that maps an item to a position, while $\pi(i)$ represents the position where item $i$ lies. Here, ${c}_{i, \pi(i)}$ denotes whether item $i$ is clicked at position $\pi(i)$ and $\gamma_i$ is a given fixed utility value for item, e.g., the utility value could be the bid price of an item for ads recommendation. 
The utility for some recommendation scenarios are total clicks, then $\gamma_i$ is set to 1. Notice that the distribution of the click probability $P(c_{i, \pi(i)} = 1)$ varies with the counterfactual permutation $\pi$. However, most of the existing works predict the estimated utility scores only based on the initial lists, overlooking the difference between the distributions of the click probability before and after reranking.

As such, the goal of utility-based reranking is to find the best permutation $\pi^*$ of candidate items after reranking for each request to maximize the total utility. The optimal utility $U^{*}$ is thereby defined as
  \begin{equation}
\begin{aligned}
     U^{*} =  \sum_{r \in \mathcal{R}} \max_{\pi}U_{\pi}(r)  
     = \sum_{r \in \mathcal{R}}\sum_{i\in L_r} P(c_{i, \pi^*(i)} = 1) \cdot \gamma_{i} ~.
\end{aligned}
\label{eq:utility-obj}
\end{equation} 

 \section{Model Framework}
This section introduces a general reranking framework, URCC, to maximize the utility in Eq.~\eqref{eq:utility}.
 Firstly, we briefly introduce the overall framework.
Then, we present the detailed design of each component.
 
  \subsection{Overall Framework}

  The architecture of our proposed URCC is shown in Figure \ref{fig:framework}. 
  First, the position-aware graph embedding is proposed to extract the item-item and item-position interaction information. For each request, we construct a fully connected item graph with the initial positions as attribute nodes and use GAT \cite{GAT} to learn item embedding. Second, taking counterfactual context and graph embeddings into account, we derive a utility evaluator, which estimates listwise utility under different contexts. Position-aware graph embedding and the evaluator are first trained together on click logs before reranker. Then, an efficient pairwise reranker is trained under the guidance of the evaluator, with the parameters of GAT and evaluator fixed. At the inference stage, only the reranker is employed to rerank items.
 In the ensuing part, we will elaborate on the details of each component.
 
 \subsection{Position-aware Graph Embedding}
 
  In the reranking stage, the item-item relationship is an essential factor where many efforts have been made to better exploit the mutual influences among items. However, we point out the \textit{relationship between items and positions} is also of cardinal significance, from the perspective of matching an item to its most suitable position to optimize the overall utility. However, less attention is paid to this aspect. Previous RNN-based and Transformer-based methods only use item-position information implicitly. To explicitly exploit item-item and item-position interaction information, we construct a fully connected item graph with the initial positions as attribute nodes for each request. As illustrated in Figure \ref{fig:framework}, node $v_i$ stands for the candidate item in a user's request with position $p_i$ as its attribute node. Mathematically, this graph can be represented as $\mathcal{G}=\{\mathcal{V}, \mathcal{E}, \mathcal{P}\}$, where node $v_i\in \mathcal{V}$ denotes item $i$ and edge $e_{i,j}\in \mathcal{E}$ indicates the connection between item $i$ and item $j$. Given that items in a list are naturally related, we build a fully connected item graph. The position attribute set $\mathcal{P}=\{\mathbf{p}_1, \mathbf{p}_2, ... ,\mathbf{p}_n\}$ contains all nodes' position information, and $\mathbf{p}_k\in \mathbb{R}^n$ is the one hot position embedding at position $k$. 
 
 Then, we aim to represent each node as a low-dimensional embedding that preserves not only item relationships but also node attribute proximity. We follow the work of graph attention network \cite{GAT}. At the $(t+1)$-th propagation step, the graph embedding layer takes as input the node feature $\mathbf H^t$ and the position attribute $\mathcal P$, where $\mathbf{H}^t=\{\mathbf{h}_1^t, \mathbf{h}_2^t, ..., \mathbf{h}_n^t\},~ \mathbf{h}_i^{t}\in \mathbb{R}^{d_t}$ denotes the node feature of item $i$ and $n$ is the number of nodes in a request, and $d_t$ is the dimension of item node's features. 
 
 In the first step, the node feature $\mathbf{H}^0$ is initialized by the item feature $\{\mathbf{x}_1, \mathbf{x}_2, ..., \mathbf{x}_n\}$. To get the normalized attention coefficients $\alpha_{i,j}^t$, a shared linear transformation function, parameterized by a weight matrix $\mathbf{W}\in \mathbb{R}^{{d_{t+1} \times (d_t + n})}$($d_{t+1}$ denotes the dimension of the new item node’s features in the output of this propagation), and the self-attention mechanism is applied to each node:
\begin{equation}
\begin{aligned}
      \mathbf{z}_i^t=\mathbf{W}[\mathbf{h}^t_i\oplus \mathbf{p}_{k_i}]~,
\end{aligned}
\label{eq:gat-0}
\end{equation}  
\begin{equation}
\begin{aligned}
      e_{i,j}^t=FNN([\mathbf{z}_i^t\oplus\mathbf{z}_j^t])~,
\end{aligned}
\label{eq:gat-1}
\end{equation} 
   \begin{equation}
\begin{aligned}
      \alpha^t_{i,j}=softmax_j(e_{i,j}^t)=\frac{e^{t}_{i,j}}{\sum_{k\in \mathcal{V}} e^{t}_{i,k}}~,
\end{aligned}
\label{eq:gat-2} 
\end{equation} 
where $FNN(\cdot)$ is a one-layer feed-forward neural network, applying a weighted matrix and LeakyReLU activate function, and $\oplus$ denotes the concatenation operation. As we use $k_i$ to represent the initial position of item $i$, $\mathbf{p}_{k_i}$ is the initial position embedding of item $i$. Finally, we use the normalized attention coefficients obtained before, followed by a nonlinear activation function $\sigma$, to output the updated node feature.
   \begin{equation}
\begin{aligned}
      \mathbf{h}^{t+1}_i = \sigma(\sum_{j\in \mathcal{V}}\alpha^t_{i,j}\mathbf{z}_j^{t}) ~.
\end{aligned}
\label{eq:gat-3} 
\end{equation}
To stabilize the performance of self-attention, we deploy the multi-head mechanism. Therefore, $M$ independent attention mechanisms are executed, and all the node features are concatenated to get the final representation of the next layer:

\begin{equation}
\begin{aligned}
      \mathbf{z}_i^{t,m}=\mathbf{W^m}[\mathbf{h}^t_i\oplus \mathbf{p}_{k_i}]~,
\end{aligned}
\label{eq:gat-4}
\end{equation}  
   \begin{equation}
\begin{aligned}
      \mathbf{h}^{t+1}_i = \vcenter{\hbox{$\mathlarger{\mathlarger{\mathlarger{\mathlarger{\oplus}}}}$}}_{m=1}^{M}\sigma\left(\sum_{j\in \mathcal{V}}\alpha_{i,j}^{t,m}\mathbf{z}_j^{t,m}\right)~,
\end{aligned}
\label{eq:gat-5} 
\end{equation}
where $\alpha_{i,j}^{t,m}$ and $\mathbf{W}^m$ denote the normalized attention coefficients and the corresponding weight matrix of linear transformation obtained in the $m$-th attention mechanism, respectively.

 After T propagation steps, we obtain the final item graph embedding for a specific request: $\mathbf{H}^T=\{\mathbf{h}^T_{1}, \mathbf{h}^T_{2}, ..., \mathbf{h}^T_{n}\}$, which capture the relationships and interactions between items and positions. 
 As such, the position-aware mutual influences between items are preserved, which will further help the training of evaluator and reranker.

 \subsection{Utility-oriented Evaluator for Counterfactual Context}
Most of the existing reranking methods directly estimate a score for each item from either human-annotated relevance labels or implicit feedback on initial lists and then places items in the decreasing order of the scores. 
However, this practice does not necessarily bring high utility for the reranked lists, since it does not take into account the \textit{gap between the context of the initial list and the reranked list}, leading to imprecise estimation. Therefore, we propose utility-oriented reranking aiming to find the list with maximal utility after reranking. However, it is still challenging to obtain the feedback for the new permutation of the reranked results.

As it is impractical to ask users to provide feedback for every possible permutation of the list, users' feedback to counterfactual context is unobtainable. We decide to design an evaluator to estimate the utility of the reranked list precisely based on observed ranking lists as well as counterfactual context. As the utility value $\gamma_i$ is a given fixed value, the goal of the evaluator is to estimate the click probability $P(c_{i, \pi(i)}=1)$, where $\pi(i)$ denotes the position of item $i$ under the permutation $\pi$. Recently, several deep models \cite{qu2018product, zhang2016deep} have been proposed to model the complex interaction from click logs, from which we can borrow some mechanisms to build our evaluator. Assuming that the click probability $P(c_{i, \pi(i)}=1)$ can be predicted by a function $g(\cdot)$ parameterized by $\bm{\theta}$,  
 \begin{equation}
\begin{aligned}
      P(c_{i, \pi(i)}=1)=g(\mathbf{x}_{i},\mathbf{h}^T_{i}, \mathbf{p}_{\pi(i)};\bm{\theta})~.
\end{aligned}
\label{eq:pred} 
\end{equation}
where $\mathbf{x}_{i}$ and $\mathbf{h}^T_{i}$ denote the feature and position-aware graph embedding of item $i$ we get in the previous part. We denote the position embedding at position $\pi(i)$ as $\mathbf{p}_{\pi(i)}$. Then the loss can be formulated as Eq. \eqref{eq:utility-loss}, where $l$ is the cross-entropy loss.

   \begin{equation}
\begin{aligned}
      \mathcal{L}(\bm{\theta})=\sum_{r\in \mathcal{R}} \sum_{i\in L_r}l(c_{i, \pi(i)}, g(\mathbf{x}_{ i},\mathbf{h}^T_{i}, \mathbf{p}_{\pi(i)}; \bm{\theta}))~.
\end{aligned}
\label{eq:utility-loss} 
\end{equation}

Specifically, users' clicks depend on the context of the items ranked both before and after the current items.
Thus, it is natural to adopt Bi-LSTM to model the user's click behaviors and capture the sequential dependencies bi-directionally in the evaluator. Formally, let $\mathbf{w}^i$, the concatenation of item feature $\mathbf{x}_{i}$, position embedding $\mathbf{p}_{\pi(i)}$, and graph embedding $\mathbf{h}_{i}^{T}$, be the input vector for the $i$-th item. The forward output state $\overrightarrow{\mathbf{q}^i}\in \mathbb{R}^{d_h}$ of size $d_h$ for the $i$-th item is computed as:
\begin{equation}
\begin{split}
\mathbf{f}^i & = \sigma(\mathbf{W}_f[\mathbf{q}^{i-1}, \mathbf{w}^i, \mathbf{c}^{i-1}] + b_f) \\
\mathbf{d}^i & = \sigma(\mathbf{W}_d[\mathbf{q}^{i-1}, \mathbf{w}^i, \mathbf{c}^{i-1}] + b_d) \\
\mathbf{c}^i & = \mathbf{f}^i * \mathbf{c}^{i-1} + \mathbf{d}^i * \tanh(\mathbf{W}_c[\mathbf{q}^{i-1}, \mathbf{w}^i] + b_c) \\
\mathbf{o}^i & =\sigma(\mathbf{W}_o[\mathbf{q}^{i-1},  \mathbf{w}^i, \mathbf{c}^{i}] + b_o) \\
\overrightarrow{\mathbf{q}^i} & = \mathbf{o}^i * \tanh(\mathbf{c}^i)
\end{split}
\label{eq:lstm} 
\end{equation}
where $\mathbf{f}, \mathbf{d}, \mathbf{o}$, and $\mathbf{c}$ are the forget gate, input gate, output gate, and cell vector with $\mathbf{W}_f$, $\mathbf{W}_d$, $\mathbf{W}_o$, and $\mathbf{W}_c$ as their trainable weight matrices, respectively, $\sigma(\cdot)$ is the logistic function, and $*$ is the element-wise product operator. Similarly, we can obtain the backward output state $\overleftarrow{\mathbf{q}^i}\in\mathbb{R}^{d_h}$. Then, we concatenate $\overrightarrow{\mathbf{q}^i}$ and $\overleftarrow{\mathbf{q}^i}$ to get the sequential representation $\mathbf{q}_{i}=[\overrightarrow{\mathbf{q}^i}\oplus \overleftarrow{\mathbf{q}^i}]$ of item $i$.

As a common and powerful technique in modeling interaction in click probability prediction task, multi-layer perception (MLP) is also integrated into our evaluator. Hence, taking the concatenation of the item features $\mathbf{x}_{i}$ and the sequential representation $\mathbf{q}_{i}$ as input, the function $g(\cdot)$ can be formalized as follows:
\begin{equation}
\begin{aligned}
     g(\mathbf{x}_{i}, \mathbf{h}^T_{i}, \mathbf{p}_{\pi(i)};\bm{\theta})=\text{MLP}(\mathbf{x}_{i}\oplus \mathbf{q}_{i})~,
\end{aligned}
\label{eq:g-funct} 
\end{equation}
where $\bm{\theta}$ denotes the union of parameters for Bi-LSTM and MLP and $\oplus$ represents the concatenation operation.

With position-aware graph embedding and Bi-LSTM modeling the context and mutual influence before and after reranking, the evaluator is able to estimate the listwise utility of any counterfactual permutation. Therefore, it is capable of providing helpful guidance for the following reranker. 

 \subsection{Reranker}
After obtaining the evaluator, one straightforward evaluation-after-reranking method might be to generate all the possible ranking lists and use the evaluator to estimate their utility. Unfortunately, given that different permutations bring different contexts and further lead to different utilities, it is requisite to get all the possible permutations of $n!$ if there are $n$ items in the initial list. Thus, this straightforward solution will have the computational complexity of $O(n!)$ at the inference stage. To avoid such high computational cost of generating all the permutations, we adopt a view of swap and use the Lambdaloss~\cite{lambdaloss} framework to \textit{adjust some improper matching towards the optimal ranking}, as shown in Figure \ref{fig:framework}.

A scoring function $\Phi(\cdot)$ parameterized by $\bm{\Theta}$ is employed to approximate the optimal matching, and then the complexity can be reduced to $O(n)$ at the inference stage. The scoring function takes the item feature $\mathbf{x}_{i}$, the obtained item graph embedding $\mathbf{h}^T_{i}$, and initial position embedding $\mathbf{p}_{k_i}$ as input, and outputs a score $s_{i}$:
 \begin{equation}
\begin{aligned}
    s_{i}=\Phi(\mathbf{x}_{i}, \mathbf{h}^T_{i}, \mathbf{p}_{k_i}; \bm{\Theta})~.
\end{aligned}
\label{eq:score-function} 
\end{equation}
After all the items' scores in a request are computed, the final list is generated by sorting the scores in descending order.

From the perspective of matching an item to its most suitable position, the initial ranking provides reasonable matching results. Therefore, we only need to exchange improper matching towards the optimal ranking by comparing the utility before and after swapping the matching between a pair of items. Hence, we adopt pairwise optimization and choose an efficient pairwise ranking framework, LambdaLoss~\cite{lambdaloss}, which is defined as
 \begin{equation}
\begin{aligned}
   \mathcal{L}_{\lambda}(r)=\sum_{i=1}^{n} \sum_{j:y_i > y_j} |\Delta NDCG(i, j)| \log(1 + e^{-\sigma(s_i-s_j)}) ~,
\end{aligned}
\label{eq:lambda-loss} 
\end{equation}
where $y_i$ and $s_i$ denote the label and predicted score of item $i$, the parameter $\sigma$ determines the shape of the sigmoid function, and $|\Delta NDCG(i, j)|$ is defined as the absolute difference between the NDCG metric before and after the two items $i$ and $j$ are swapped.
Notice that the original LambdaLoss framework aims to optimize the NDCG metric while our goal is to maximize the total utility. Therefore, we derive an unbiased estimate of the listwise utility $U_{\pi}(r)$ in Eq.~\eqref{eq:utility} to replace the original NDCG. For any permutation $\pi$ in request $r$, our designed utility metric is
 \begin{equation}
\begin{aligned}
   u_{\pi}(r)=\sum_{i=1}^{n} c_{i, k_{i}} \cdot \frac{P(c_{i, \pi(i)}=1)}{P(c_{i, k_{ i}}=1)}\cdot \gamma_{i}~,
\end{aligned}
\label{eq:utility-metrics} 
\end{equation}
where $P(c_{i, \pi(i)}=1)$ and $P(c_{i, k_{i}}=1)$ denote the estimated click probability of item $i$ displayed in the position of permutation $\pi$ and  historical logs, which is estimated by the utility-oriented evaluator. $\gamma_i$ is a given fixed utility value for item $i$. The utility metric can be proved unbiased by showing the expectation of $u_{\pi}$(r) is equivalent to $U_{\pi}$(r) , as
 \begin{equation}
\begin{aligned}
   \mathbb{E}[u_{\pi}(r)]  
   & =\mathbb{E} \Big[\sum_{i=1}^{n} c_{i, k_{i}} \cdot \frac{P(c_{i, \pi(i)}=1)}{P(c_{i, k_{i}}=1)}\cdot \gamma_{i} \Big] \\
   & = \sum_{i=1}^{n} P(c_{i, k_{i}}=1) \cdot \frac{P(c_{i, \pi(i)}=1)}{P(c_{i, k_{ i}}=1)}\cdot \gamma_{i} \\
   & = \sum_{i=1}^{n} P(c_{i,\pi(i)}=1)\cdot \gamma_{i} \\
   & = U_{\pi}(r)~.
\end{aligned}
\label{eq:utility-prove} 
\end{equation}


Then, similar to $|\Delta NDCG(i, j)|$, we sample item pairs from the ranking list provided by the initial ranker, and then calculate the difference between the listwise utility metric before and after the two items $i$ and $j$ are swapped:
 \begin{equation}
\begin{aligned}
   \Delta Utility(i, j) = u_{\pi'}(r) - u_{\pi}(r)~,
\end{aligned}
\label{eq:diff-utility} 
\end{equation}
where $\pi$ is the original permutation of the initial list and $\pi'$ denotes the new permutation after item $i$ and $j$ are swapped. 
With $\Delta Utility(i, j)$ as the weight for each pair, we derive the loss function for training the reranker:
 \begin{equation}
\begin{aligned}
  \mathcal{L}_{\lambda}(r; \bm{\Theta})=\sum_{i=1}^{n} \sum_{j:k_{i} > k_{j}} \Delta Utility(i, j) \log(1 + e^{-\sigma(s_{i}-s_{j})}) ~.
\end{aligned}
\label{eq:lambda-utility-loss} 
\end{equation}

\section{Experiments}
In this section, we first compare our proposed CRUM model with the state-of-the-art reranking algorithms on two public datasets and a real-world industrial dataset. Secondly, we investigate the impact of different components and hyper-parameters of CRUM. Finally, we study the reranking performance of CRUM by varying the quality of the initial ranking lists. 

\subsection{Experimental Setup}
\subsubsection{Datasets.}
Our experiments are conducted on two public learning-to-rank benchmark datasets, including Yahoo! LETOR set 1
\footnote{https://webscope.sandbox.yahoo.com}
and Microsoft MSLR-WEB10K
\footnote{https://www.microsoft.com/en-us/research/project/mslr/}
, and a large scale proprietary dataset obtained from a real-world App Store. 
\begin{itemize}
    \item \textbf{Yahoo! LETOR set 1} (Yahoo for short) is used in Yahoo! Learning-to-Rank Challenge, consisting of 700 features normalized in $[0,1]$ extracted from query-document pairs.
    
    \item \textbf{Microsoft MSLR-WEB10K} (MSLR for short) is a large-scale dataset released by Microsoft Research in May 2010. It is composed of 10,000 queries and 1,200,193 documents with 136 features extracted from query-document pairs. 
    
    \item \textbf{App Store} contains user click logs from Jan 8, 2021 to Jan 31, 2021 from a mainstream industrial App Store, with 27,928,214 users, 398,053,272 items. Each item has 30 features, including features of the item itself, user features, and context features. 
\end{itemize}

\subsubsection{Initial ranker and baselines.}
To generate the initial ranking lists, we select three representative learning-to-rank algorithms, including DNN, SVMrank, and LambdaMART. Those three algorithms use pointwise, pairwise, and listwise loss, respectively. 
\begin{itemize}
    \item \textbf{DNN} \cite{youtube} applies MLP to model the relationship between labels and features of items with pointwise loss.
    \item \textbf{SVMrank} \cite{svmrank} is a classic pairwise learning-to-rank model built upon the SVM algorithm.
    \item \textbf{LambdaMART} \cite{burges2010ranknet} is the state-of-the-art listwise learning-to-rank algorithm, which 
    optimizes NDCG directly.
\end{itemize}

We list the reranking solutions for our empirical comparisons below.
As for IRGPR and PRS we mentioned before, IRGPR \cite{irgpr} demands an item relationship graph and PRS requires multiple user behavioral logs, like browsing, favoring, and buying, making them hard to implement in the datasets we use. 
\begin{itemize}

    \item \textbf{Seq2Slate} \cite{seq2slate} uses pointer network to sequentially encode previously selected items and uses decoder to predict the next one.
    \item \textbf{DLCM} \cite{dlcm} first applies GRU, which encodes top-ranking items to learn a local context embedding, and then combines it and the original feature to rerank the top results.
    \item \textbf{PRM} \cite{prm} employs self-attention mechanism to model the mutual influence between items and users' preferences.
    \item \textbf{SetRank} \cite{setrank} employs a stack of  multi-head self-attention blocks to learn a permutation-invariant ranking model.
\end{itemize}

\subsubsection{Click data generation.}
Given that Yahoo and MSLR are datasets with human-annotated relevance labels, synthetic click data is necessary to simulate user click behavior. Amongst the baselines, DLCM and SetRank directly use human-annotated relevance labels. The click generation adopted by PRM only takes into account the relevance labels and position decay, ignoring the high-order interaction between clicks. Thus, we mainly follow Seq2Slate \cite{seq2slate} to generate synthetic clicks for the two datasets. Firstly, we convert the original ratings (0 to 4) to binary labels with a threshold $T_b=1$ (relevant: \{2, 3, 4\}, irrelevant: \{0, 1\}) to obtain relevance probability $rel(r, i)$. Secondly, a user observes each item with a position decaying probability $1/pos(i)^{\eta}$, where $pos(i)$ is the ranking position of item $i$ and $\eta$ is a decay parameter. Here we set $\eta$ to 0.7.  Then, when observing an item, the user will click if it is similar to previous clicked items. The similarity probability introduces high-order interaction between clicks. Here, we modify the original deterministic similarity in Seq2Slate to a more reasonable similarity probability.  If there exists an item clicked before, the cosine similarity between the currently browsing item and the previously clicked item will serve as the similarity probability. If not, the similarity probability will be set to 1. Finally, the click probability is the product of relevance, position decaying, and similarity probability.

\subsubsection{Evaluation metrics} 
For Yahoo and MSLR, all baselines and our model CRUM is evaluated in terms of the relevance-based metrics \textit{MAP} and \textit{nDCG} \cite{ndcg}, and the utility-based metrics \textit{\# Click} and \textit{CTR}. Here, \textit{\# Click} and \textit{CTR} denote \textit{number of clicks per list} and \textit{click probability per item} in the reranked lists evaluated by the oracle click model used in click data generation section. Relevance-based metrics are evaluated with binarized relevance labels. 
For the proprietary dataset App Store, $nDCG@K$ and $Revenue@K$ are adopted as relevance-based and utility-based metrics, respectively. These two metrics are valuated based on click logs, following PRM \cite{prm}. Mathematically, $Revenue@K$ is defined as the average of  the expected revenue at top-$K$ positions in Eq. \eqref{eq:revenue}
   \begin{equation}
\begin{aligned}
      Revenue@K=\frac{1}{|\mathcal{R}|}\sum_{r\in \mathcal{R}}\sum_{k=1}^K b_{\omega(k)}\cdot c_{\omega(k)}^h~,
\end{aligned}
\label{eq:revenue} 
\end{equation} 
where $\omega$ denotes the ordered list of items and $\omega(k)$ is the item reranked at position $k$. $b_{\omega(k)}$ is the bid price of item $\omega(k)$, and $c_{\omega(k)}^h$ denotes whether the item is clicked in the click logs.

\subsubsection{Reproducibility.} For public datasets Yahoo and MSLR, we implement our model and baselines in Tensorflow 1.9.0. This implementation of our model is available for reviewers\footnote{Code for our experiments is available at https://bit.ly/3ot2Heg} and will be publicly available upon the acceptance of this work. We employ Xavier method \cite{xavier} to initialize model parameters and use Adam \cite{adam} as optimizer. The maximum number of positions is set to 10.
Firstly, we train the graph embedding and evaluator together. There are two layers in GAT, and the embedding size of GAT and Bi-LSTM are both 64.  The architecture of MLP in evaluator is [1024, 512, 128, 64]. The learning rate of the optimizer is 0.0003, and the batch size is set to 128. Then, the reranker is trained with the parameters of GAT and evaluator fixed. The architecture of MLP in reranker is the same as the evaluator, and the number of sampled pairs for each list is 10. The learning rate and batch size are set to 0.00001 and 128. For baselines, all hyper-parameters and initialization strategies follow the suggestion from their paper or are tuned on the validation sets. 

Due to the change of environment and labels, the implementation on the private dataset has some differences. Most of the settings remain the same with public datasets, and only the modified ones are listed as follows. This model is implemented in Tensorflow 1.4.0. The maximum number of positions is set to 30. Instead of MLP, DCN is employed in reranker with three cross-layer and [1025, 512, 256, 128] as the architecture of the deep part. In the evaluator's training, the learning rate is 0.0005, and the batch size is set to 100. In reranker, the number of sampled pairs for each list is 15.

\begin{table*}[htbp]
\newcommand{\tabincell}[2]{\begin{tabular}{@{}#1@{}}#2\end{tabular}}
	\centering
	\footnotesize
	\caption{Comparison of different reranking models on two benchmark datasets.
	}
	\vspace{-2pt}
	\scalebox{1.12}{
	\begin{tabular}{ c| c | c || c | c | c | c  |c |c |c |c | c | c} 
		\hline
        \multicolumn{1}{c|}{\multirow{3}{*}{Initial Ranker}} &
        \multicolumn{2}{c||}{\multirow{3}{*}{\tabincell{c}{Reranking\\Model}}} & \multicolumn{5}{c|}{Microsoft MSLR-WEB10K} & \multicolumn{5}{c}{Yahoo! LETOR set 1}  \\ \cline{4-13} 
        \multicolumn{1}{c|}{} & \multicolumn{2}{c||}{} & \multicolumn{3}{c|}{Relevance-based} & \multicolumn{2}{c |}{Utility-based} & \multicolumn{3}{c|}{Relevance-based} & \multicolumn{2}{c}{Utility-based}  \\ \cline{4-13} 
		\multicolumn{1}{c|}{} & \multicolumn{2}{c||}{} & MAP & nDCG@5 & nDCG@10 & \# Click & CTR  & MAP & nDCG@5 & nDCG@10 & \# Click & CTR  \\\hline
    \multirow{6}{*}{DNN}
        &\multicolumn{2}{c||}{None}  & 0.5338  & 0.4970 & 0.6775 & 1.7180 & 0.1744 & 0.6935  & 0.6704 & 0.8012 & 2.8978 & 0.3151\\\cline{2-3}
		&\multicolumn{2}{c||}{DLCM}  & 0.5397 & 0.5052 & 0.6818 & 1.7730 & 0.1768  & 0.7141  & 0.6958 & 0.8170 & 2.9810 & 0.3267\\ \cline{2-3}
		&\multicolumn{2}{c||}{Seq2Slate}  & 0.5470  & 0.5170 & 0.6888 & 1.8046 & 0.1789  & 0.7267 & 0.7006 & 0.8167 & 3.0590 & 0.3327\\\cline{2-3}
		&\multicolumn{2}{c||}{PRM}  & 0.5531  & 0.5232 & 0.6945 & 1.7767 & 0.1808  & 0.7265  & 0.7001 & 0.8178 & 3.0642 & 0.3332\\\cline{2-3}
		&\multicolumn{2}{c||}{SetRank} & 0.5404  & 0.5095 & 0.6820 & 1.7563 & 0.1770 & 0.7160  & 0.6973 & 0.8176 & 3.0111 & 0.3275 \\\cline{2-3}
		&\multicolumn{2}{c||}{CRUM} & \textbf{0.5827*} & \textbf{0.5557*} & \textbf{0.7161*} & \textbf{1.8712*} & \textbf{0.1860*} & \textbf{0.7385*} & \textbf{0.7216*} & \textbf{0.8343*} & \textbf{3.0915*} & \textbf{0.3409*} \\\hline
	    \multirow{6}{*}{SVMRank}	
		&\multicolumn{2}{c||}{None} & 0.5279  & 0.4982 &0.6735 &1.7156 &0.1720 & 0.6790 & 0.6518 &0.7914 & 2.7957 &0.3032\\\cline{2-3}
		&\multicolumn{2}{c||}{DLCM} & 0.5365  & 0.5032& 0.6957 & 1.7816 & 0.1755 & 0.7102 & 0.6898 & 0.8143 & 2.9875 & 0.3219 \\ \cline{2-3}
		&\multicolumn{2}{c||}{Seq2Slate} & 0.5494  & 0.5177 & 0.6915 & 1.7857 & 0.1788 & 0.7166 & 0.6974 & 0.8192 & 2.9791 & 0.3245  \\ \cline{2-3}
		&\multicolumn{2}{c||}{PRM} & 0.5540  & 0.5224 & 0.6957 & 1.8391 & 0.1798  & 0.7187 & 0.7001 & 0.8206 & 2.9789 & 0.3252 \\ \cline{2-3}
		&\multicolumn{2}{c||}{SetRank} & 0.5377  & 0.5059 & 0.6805 & 1.8126 & 0.1759 & 0.7090 & 0.6885 & 0.8132 & 2.9715 & 0.3214 \\ \cline{2-3}
		&\multicolumn{2}{c||}{CRUM} & \textbf{0.5806*}  & \textbf{0.5578*} & \textbf{0.7142*} & \textbf{1.8599*} & \textbf{0.1853*} & \textbf{0.7353*} & \textbf{0.7192*} & \textbf{0.8328*} & \textbf{3.0801*} & \textbf{0.3379*} \\ \hline
        \multirow{6}{*}{LambdaMART}
        &\multicolumn{2}{c||}{None}  & 0.5465  & 0.5124 & 0.6866 & 1.7726 & 0.1769  & 0.7140 & 0.6925 & 0.8188 & 2.9791 & 0.3239 \\ \cline{2-3}
		&\multicolumn{2}{c||}{DLCM}  & 0.5506  & 0.5190 & 0.6897 & 1.8003 & 0.1789  & 0.7260 & 0.7075 & 0.8267 & 3.0512 & 0.3318 \\ \cline{2-3}
		&\multicolumn{2}{c||}{Seq2Slate}  & 0.5686  & 0.5434 & 0.7050 & 1.8028 &0.1826 & 0.7331  & 0.7167 & 0.8318 & 3.0570 &0.3344 \\ \cline{2-3}
		&\multicolumn{2}{c||}{PRM}  & 0.5699  & 0.5438 & 0.7059 & 1.8266 & 0.1828  & 0.7339 & 0.7163 & 0.8327 & 3.0661 & 0.3351 \\ \cline{2-3}
		&\multicolumn{2}{c||}{SetRank} & 0.5531  & 0.5235 & 0.6911 & 1.7922 & 0.1792 & 0.7282 & 0.7084 & 0.8276 & 3.0539 & 0.3329 \\ \cline{2-3}
		&\multicolumn{2}{c||}{CRUM}  & \textbf{0.5855*} & \textbf{0.5629*} & \textbf{0.7165*} & \textbf{1.8688*} & \textbf{0.1857*} & \textbf{0.7360*} & \textbf{0.7197*} & \textbf{0.8335*} & \textbf{3.1116*} & \textbf{0.3392*}
		\\ \hline
	\end{tabular}
	}
	\vspace{0pt}
	\footnotesize \flushleft $*$ denotes statistically significant improvement (measured by t-test with $p$-value$<$0.05) over all baselines.
	\label{tab:overall}
\end{table*}

\begin{table*}[htbp]
\newcommand{\tabincell}[2]{\begin{tabular}{@{}#1@{}}#2\end{tabular}}
	\centering
	\caption{Comparison of different reranking models on the real-world App Store dataset. 
	}
	\vspace{-5pt}
	\resizebox{0.92\textwidth}{!}{
	\begin{tabular}{ c| c | c || c | c | c | c | c | c| c|c} 
		\hline
		 \multicolumn{3}{c||}{Reranking Model}& nDCG@3  & nDCG@5 & nDCG@10 & nDCG@20 & Revenue@3 & Revenue@5 & Revenue@10 &Revenue @20 \\\hline
		 
		\multicolumn{3}{c||}{initial} & 0.3045 & 0.3646 & 0.4281 & 0.4848 & 2.461 & 3.313 & 4.292 & 5.304\\\hline
		\multicolumn{3}{c||}{DLCM} & 0.2943 & 0.3631 & 0.4323 & 0.4839 &  2.634 & 3.478 & 4.461 & 5.393  \\ \hline
		\multicolumn{3}{c||}{Seq2Slate}& 0.2861 & 0.3541 & 0.4223 & 0.4776 & 2.640 & 3.468 & 4.466 & 5.390  \\\hline
		\multicolumn{3}{c||}{PRM} & 0.3020 & 0.3648 & 0.4342 & 0.4883 & 2.676 & 3.478 & 4.484 & 5.401   \\\hline
		\multicolumn{3}{c||}{SetRank}  & 0.3077 & 0.3842 & 0.4599 & 0.4993 & 2.491  & 3.380  & 4.421 & 5.353     \\\hline
		\multicolumn{3}{c||}{CRUM} & \textbf{0.3477*} & \textbf{0.4127*} & \textbf{0.4802*} & \textbf{0.5255*} & \textbf{2.688*} & \textbf{3.503*} & \textbf{4.494*} & \textbf{5.405}  \\ \hline
	\end{tabular}
	}
	\vspace{-0pt}
	\footnotesize \flushleft $*$ denotes statistically significant improvement (measured by t-test with $p$-value$<$0.05) over all baselines.
	\vspace{0pt}
	\label{tab:huawei}
\end{table*}
 
\subsection{Overall Performance}
\subsubsection{Benchmark datasets.}
 The overall performance on the two benchmark datasets, Yahoo and MSLR, is reported in Table \ref{tab:overall}, from which we have several important observations.

Firstly, our proposed CRUM significantly and consistently outperforms the state-of-the-art approaches w.r.t. both relevance-based and utility-based metrics under three initial rankers on both datasets. Taking MSLR as an example, CRUM improves over the best baseline PRM w.r.t utility-based metric \textit{CTR} by 2.88\%, 3.06\%, and 1.59\% on DNN, SVMRank, and LambdaMART, respectively. In terms of the relevance-based metric \textit{MAP}, CRUM also achieves 5.35\%, 4.80\%, 2.74\% improvement over the best baseline on three initial rankers. This demonstrates the effectiveness of evaluation-after-reranking method and the modeling of counterfactual context. Besides, the results also suggest utility-based and relevance-based metrics are correlated on the two datasets, i.e., a method performs well on utility-based and relevance-based metrics simultaneously.

Secondly, the initial rankers with different forms of loss function affect the performance of CRUM in various ways. 
In Table \ref{tab:overall}, LambdaMART performs better than DNN, and DNN performs better than SVMRank. 
Although 
the pairwise method usually is  more significant than the pointwise method, the deep-learning-based method DNN outperforms SVMRank due to the superior expressive ability of the deep network. 
Nevertheless, the result of CRUM on these three initial rankers is not consistent with the performance of the initial rankers, e.g., CRUM on DNN achieves better results than that on LambdaMART. The reason may be that listwise information is critical in reranking. DNN uses a pointwise loss function without listwise interaction information, and thus has a considerable improvement space. However, LambdaMART has already incorporated listwise information in loss function,
which limits its room for improvement, so that the performance of CRUM on LambdaMART is inferior to that on DNN.

Finally,  we observe that RNN-based algorithms achieve relatively poorer performance than Transformer-based ones in most cases. Transformer-based method PRM yields the best results among all baselines, demonstrating the effectiveness of self-attention to model mutual influence between any items. Though employing self-attention, SetRank fails to surpass the RNN-based model Seq2slate. One possible reason is that this approach is more suitable for human-annotated relevance labels, as it adopts in its original setting \cite{setrank}.

\begin{table*}[htbp]
	\centering
	\caption{Comparison of CRUM and its variants on two benchmark datasets. 
	}
	\vspace{-2pt}
	\scalebox{0.92}{
	\begin{tabular}{ c| c | c || c | c | c | c  |c |c |c |c|c|c} 
		\hline
	\multicolumn{1}{c|}{\multirow{3}{*}{Initial Ranker}} &
        \multicolumn{2}{c||}{\multirow{3}{*}{Variants}} & \multicolumn{5}{c|}{Microsoft MSLR-WEB10K} & \multicolumn{5}{c}{Yahoo! LETOR set 1}  \\ \cline{4-13} 
        \multicolumn{1}{c|}{} & \multicolumn{2}{c||}{} & \multicolumn{3}{c|}{Relevance-based} & \multicolumn{2}{c |}{Utility-based} & \multicolumn{3}{c|}{Relevance-based} & \multicolumn{2}{c}{Utility-based}  \\ \cline{4-13} 
		\multicolumn{1}{c|}{} & \multicolumn{2}{c||}{} & MAP & nDCG@5 & nDCG@10 & \# Click & CTR  & MAP & nDCG@5 & nDCG@10 & \# Click & CTR  \\\hline
		
		\multirow{3}{*}{DNN}
		&\multicolumn{2}{c||}{CRUM(-BL)}  & 0.5604  & 0.5313 & 0.6986 & 1.8011 & 0.1819& 0.7245  & 0.7064 & 0.8247 & 3.0500 & 0.3356 \\\cline{2-3}
		&\multicolumn{2}{c||}{CRUM(-GAT)}  & 0.5586  & 0.5289 & 0.6979 & 1.8460 & 0.1821 & 0.7337  & 0.7157 & 0.8305 & 3.0497 & 0.3372 \\\cline{2-3}
		&\multicolumn{2}{c||}{CRUM(-GE)} & 0.5654 &0.5362 & 0.7033 & 1.8297 & 0.1834 & 0.7385  &0.7233 & 0.8347 & 3.0875 & 0.3374\\\cline{2-3}
		&\multicolumn{2}{c||}{CRUM} & \textbf{0.5827*}  & \textbf{0.5557*} & \textbf{0.7161*} & \textbf{1.8712*} & \textbf{0.1860*}& \textbf{0.7385*} & \textbf{0.7216*} & \textbf{0.8343*} & \textbf{3.0915*} & \textbf{0.3409*}  
		\\\hline
		
	    \multirow{3}{*}{SVMRank}
		&\multicolumn{2}{c||}{CRUM(-BL)}  & 0.5546  & 0.5250 & 0.6954 & 1.8044 & 0.1810 & 0.7206  & 0.7030 & 0.8219 & 3.0439 & 0.3334 \\\cline{2-3}
		&\multicolumn{2}{c||}{CRUM(-GAT)}   & 0.5522 & 0.5221 & 0.6927 & 1.8085 & 0.1801 & 0.7317  & 0.7149 & 0.8296 & 3.0781 & 0.3365 \\\cline{2-3}
		&\multicolumn{2}{c||}{CRUM(-GE)}  & 0.5714  & 0.5472 & 0.7070 & 1.8411 & 0.1828 & 0.7330 &  0.7171 & 0.8305 & 3.0585 & 0.3366 \\\cline{2-3}
	   &\multicolumn{2}{c||}{CRUM}& \textbf{0.5806*}  & \textbf{0.5578*} & \textbf{0.7142*} & \textbf{1.8599*} & \textbf{0.1853*} &  \textbf{0.7353*}  & \textbf{0.7192*} & \textbf{0.8328*} & \textbf{3.0801*} & \textbf{0.3379*}  \\ \hline
	
		\multirow{3}{*}{LambdaMART}
		&\multicolumn{2}{c||}{CRUM(-BL)} & 0.5593  & 0.5305 & 0.6983 & 1.7995 & 0.1815 & 0.7248 & 0.7083 & 0.8246 & 3.0529 & 0.3358 \\\cline{2-3}
		&\multicolumn{2}{c||}{CRUM(-GAT)}   & 0.5599 & 0.5274 & 0.6985 & 1.8216 & 0.1814   & 0.7327 & 0.7155 & 0.8307 & 3.0673 & 0.3370 \\\cline{2-3}
		&\multicolumn{2}{c||}{CRUM(-GE)} & 0.5637  & 0.5340 &  0.7019 & 1.8471 & 0.1832  & 0.7335 & 0.7163 & 0.8311 & 3.0830 & 0.3370 \\\cline{2-3}
		&\multicolumn{2}{c||}{CRUM}  & \textbf{0.5855*} & \textbf{0.5629*} & \textbf{0.7165*} & \textbf{1.8688*} & \textbf{0.1857*} & \textbf{0.7360*} & \textbf{0.7197*} & \textbf{0.8335*} & \textbf{3.1116*} & \textbf{0.3392*} 
		\\ \hline
	\end{tabular}
	}
	\vspace{0pt}
	\footnotesize \flushleft $*$ denotes statistically significant improvement (measured by t-test with $p$-value$<$0.05) over all variants.
	\vspace{0pt}
	\label{tab:ablation}
\end{table*}

\subsubsection{Proprietary dataset.}
In order to verify the effectiveness of our proposed model on real-world click-through data, we also evaluate CRUM and the baseline models on a proprietary click-through dataset from a mainstream industrial App Store. The objective of the industrial platform is to optimize revenue. Thus we modify the proposed CRUM and the baselines accordingly by predicting the revenue (CTR * bid), instead of clicks. 


From the overall performance shown in Table \ref{tab:huawei}, we have the following observations. Firstly, CRUM consistently yields the best performance in all cases. For example, CRUM improves over the best baseline in relevance-based metrics, SetRank, by 12.99\%, 7.42\% in \textit{nDCG@3}, \textit{nDCG@5}. In terms of utility-based metrics, CRUM outperforms the best baseline PRM by 0.45\%, 0.72\% in \textit{Revenue@3}, \textit{Revenue@5}. These results demonstrate the superiority of our approach over the baselines in optimizing both relevance and utility 
via modeling counterfactual context. Secondly, unlike the semi-synthetic experiments on public datasets, there is an inconsistency between relevance-based and utility-based metrics due to the consideration of the bid prices.
When optimizing the revenue, most baselines achieve poor performance in relevance-based metrics, and some baselines (e.g., DLCM and Seq2Slate) even perform worse than the initial ranker. 
Nevertheless, although there is a trade-off between the two metrics, our model manages to find a balance to achieve much improvement in both metrics.

\subsection{In-depth Analysis}

\subsubsection{Ablation study. }
To better understand the impact of each component of CRUM, we design three variants of CRUM, which is listed as follows:
\begin{itemize}
    \item CRUM(-BL) removes Bi-LSTM from the evaluator.
    \item CRUM(-GAT) removes the graph embedding from both reranker and evaluator.
    \item CRUM(-GE) removes the graph embedding used in the reranker without modifying the evaluator.
\end{itemize}
The comparison of these three variants and original CRUM on Yahoo and MSLR datasets is shown in Table \ref{tab:ablation}. After removing each component, the performance of CRUM has declined to a certain extent w.r.t. all metrics, which demonstrates the effectiveness of Bi-LSTM and graph in leveraging counterfactual context. CRUM(-BL) roughly performs worst, proving the importance of listwise interaction. Comparing to CRUM, the performance of CRUM(-GE) and CRUM(-GAT) also have declined and removing GAT depresses the performance even more. This indicates the position-aware graph embedding not only helps the evaluator to model the counterfactual context, but also improves the performance of the reranker.

\subsubsection{Hyper-parameter study}

\begin{figure}[hbt] 
	\vspace{-0pt}
	\centering
	\includegraphics[width=0.235\textwidth]{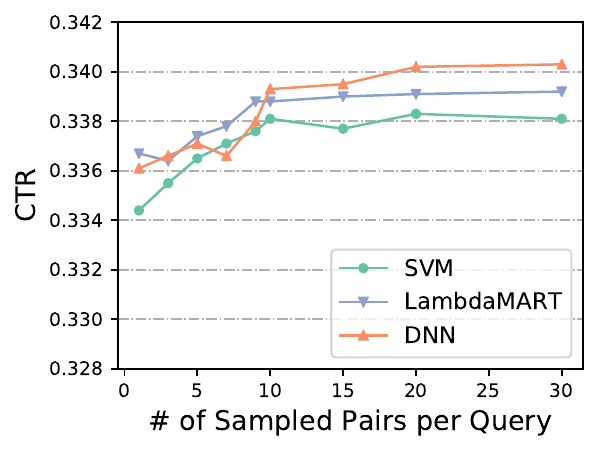}
	\includegraphics[width=0.235\textwidth]{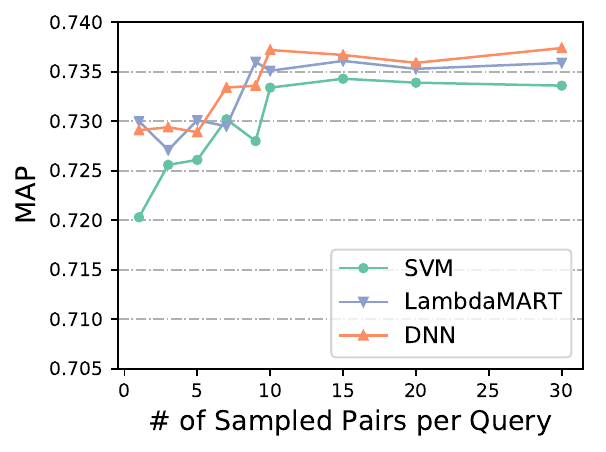}
	\vspace{-18pt}
	\caption{Impact of sampled pairs on CTR and MAP. }
	\label{fig:sample}
	\vspace{-5pt}
\end{figure}

Practically, we notice that a hyper-parameter, the number of sampled pairs of user's requests used in training, influences the final results. Thus, we conduct grid-search experiments on Yahoo. We fix all the other hyper-parameters and tune the number of sampled pairs per list from 1 to 30. Then, we visualize the change of a relevance-based metric (\textit{MAP}) and a utility-based metric (\textit{CTR}) in Figure \ref{fig:sample}.

We observe both \textit{CTR} and \textit{MAP} improve sharply from 1 to 10 and then become stable from 10 to 30. Though sampling 30 pairs per list slightly outperforms others, more samples require more space for training the model, so we set the number to 10 in our experiments. Besides, we also notice that the convergence speed is faster with more samples during the training process. 

\subsubsection{Reranking in bad cases} 
\begin{figure}[hbt]
	\vspace{-0pt}
	\centering
	\includegraphics[width=0.235\textwidth]{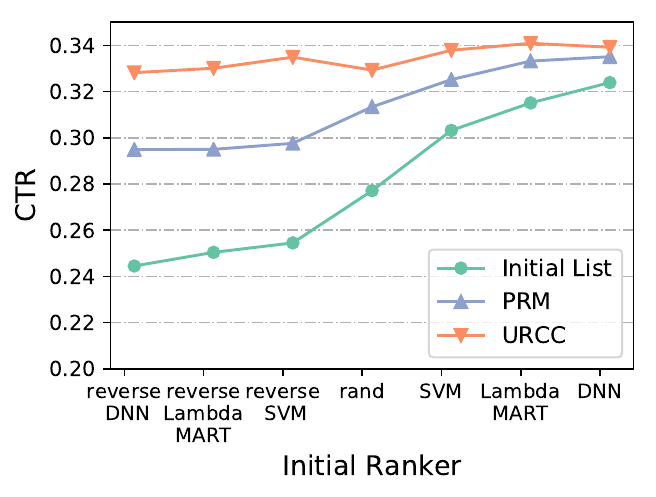}
	\includegraphics[width=0.235\textwidth]{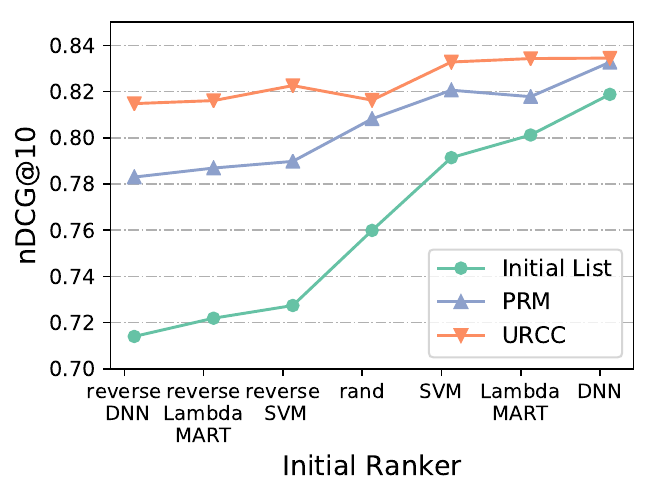}
	\vspace{-18pt}
	\caption{Impact of bad initial ranker.}
	\label{fig:badcase}
	\vspace{0pt}
\end{figure}

CRUM models the counterfactual context rather than the context given by the initial ranker, and initial ranking only serves as the starting point for the swap of the reranker. Therefore, intuitively, our model does not rely too much on the initial ranker.
To verify this, we explore reranking models with different qualities of the initial ranking lists on Yahoo in Figure \ref{fig:badcase}. Here, \textit{rand} means that the initial ranking is randomly generated. In addition, the initial ranking lists given by the original DNN, SVMRank, and LambdaMART are inverted to obtain the reverse DNN, reverse SVMRank, and reverse LambdaMART, which all perform worse than \textit{rand} ranker. Under these poor rankings, we compare our method CRUM to the strongest baseline PRM.

As illustrated in Figure \ref{fig:badcase}, we can see that as the ranking gets worse, the performance of CRUM and PRM both have a tendency to decline. 
Yet CRUM achieves more stable performance than PRM, which indicates that CRUM does rely less on the initial ranking. During training, we also observe that although initialization affects the convergence speed, it usually converges a little faster when trained with a better initial ranker. It is probably because a better initial ranker provides a better arrangement so that it takes less time for swapping optimization to achieve the optimal ranking. Besides, our approach achieves a relatively poor performance in random cases. 
One possible reason is that completely random matching is worse than inverse sorting because the matching is unordered and requires more swapping to reach the optimal position.

\section{conclusion}
In this paper, we highlight the necessity of leveraging counterfactual context to evaluate utility after reranking and address how to use such information via a general evaluation-after-reranking solution. To avoid exponential candidate lists, we propose a novel utility-oriented reranking framework, CRUM, consisting of position-aware graph embedding, utility-oriented evaluator, and pairwise reranker. The utility-oriented evaluator is designed to estimate the listwise utility via the counterfactual context modeling, and the pairwise reranker find the most suitable position for each item after reranking efficiently. Extensive experiments on two widely used public datasets and a proprietary real-world industrial dataset demonstrate CRUM's effectiveness, compared to state-of-the-art models w.r.t both relevance-based and utility-based metrics.  

\balance
\bibliographystyle{ACM-Reference-Format}
\bibliography{URCC}

\end{document}